\documentclass[12pt]{article}
\usepackage[T1]{fontenc}
\usepackage{sansmath}
\sansmath
\usepackage{helvet}

\usepackage{geometry}
\usepackage{enumitem,amssymb,amsmath}
\usepackage{graphics,graphicx}
\usepackage{sidecap}
\usepackage{multicol}
\usepackage{natbib}
\usepackage{xcolor}
\usepackage{hyperref}
\hypersetup{colorlinks,linkcolor={black},citecolor={blue},urlcolor={red}}  

\definecolor{darkblue}{HTML}{0000A0}

\usepackage{mdwlist}
\usepackage{lastpage}
\usepackage{fancyhdr}

\usepackage[font={footnotesize}]{caption}

\newcommand\chandra{{\it Chandra}}

\newenvironment{tightcenter}{%
	\setlength\topsep{0pt}
	\setlength\parskip{0pt}
	\begin{center}
}{%
 	\end{center}
}

\makeatletter
\renewcommand{\section}{\@startsection%
{section}{1}{0mm}{-\baselineskip}%
{0.5\baselineskip}{\normalfont\large\bfseries}}%
\makeatother

\fancypagestyle{author}{\fancyhf{}\fancyfoot[R]{{\footnotesize$^{*}$Corresponding Authors:\\ bsnios@cfa.harfvard.edu, w.dunn@ucl.ac.uk}}}

\begin{document}
\thispagestyle{author}

\begin{tightcenter}
\vspace*{\fill}
{\Huge \textcolor{darkblue}{X-rays Studies of the Solar System}\\} 
{\small (Thematic Area: Planetary Systems)}
\vspace{1em}
\end{tightcenter}

\begin{sloppypar}
{\large
\noindent Bradford Snios$^{1,*}$, William R.  Dunn$^{1,2,*}$, Carey M. Lisse$^{3}$, Graziella Branduardi-Raymont$^{2}$, Konrad Dennerl$^{4}$, Anil Bhardwaj$^{5}$, G. Randall Gladstone$^{6}$, Susan Nulsen$^{1}$, Dennis Bodewits$^{7}$, Caitriona M. Jackman$^{8}$, Juli{\'a}n D. Alvarado-G{\'o}mez$^{1}$, Emma J. Bunce$^{9}$, Michael R. Combi$^{10}$, Thomas E. Cravens$^{11}$, Renata S. Cumbee$^{12}$, Jeremy J. Drake$^{1}$, Ronald F.  Elsner$^{13}$, Denis Grodent$^{14}$, Jae Sub Hong$^{1}$, Vasili Kharchenko$^{1,15}$, Ralph P. Kraft$^{1}$, Joan P. Marler$^{16}$, Sofia P. Moschou$^{1}$, Patrick D. Mullen$^{17}$, Scott J. Wolk$^{1}$,  Zhonghua Yao$^{14}$
}
\end{sloppypar}

\vspace{1em}
\begin{tightcenter}
{\footnotesize
$1$. Center for Astrophysics | Harvard \& Smithsonian, Cambridge, MA, USA \\
$2$. University College London, London, UK\\ 
$3$. Johns Hopkins University Applied Physics Laboratory, Laurel, MD, USA\\
$4$. Max-Planck-Institut f\"ur extraterrestrische Physik, Garching, Germany\\
$5$. Physical Research Laboratory, Ahmedabad, India\\
$6$. Southwest Research Institute, San Antonio, TX, USA\\
$7$. Auburn University, Auburn, AL\\
$8$. University of Southampton, Southampton, UK\\
$9$. University of Leicester, Leicester, UK\\
$10$. University of Michigan, Ann Arbor, MI, USA\\
$11$. University of Kansas, Lawrence, KS, USA\\
$12$. NASA Goddard Space Flight Center, Greenbelt, MD, USA\\
$13$. NASA Marshall Space Flight Center, Huntsville, AL, USA\\
$14$. Universit{\'e} de Li{\`e}ge, Li{\`e}ge, Belgium\\
$15$. University of Connecticut, Storrs, CT, USA\\ 
$16$. Clemson University, Clemson, SC, USA\\
$17$. University of Illinois Urbana-Champaign, Champaign, IL, USA 
}
\vspace{1em}

\end{tightcenter}

\noindent{\it X-ray observatories contribute fundamental advances in Solar System studies by probing Sun-object interactions, developing planet and satellite surface composition maps, probing global magnetospheric dynamics, and tracking astrochemical reactions. Despite these crucial results, the technological limitations of current X-ray instruments hinder the overall scope and impact for broader scientific application of X-ray observations both now and in the coming decade. Implementation of modern advances in X-ray optics will provide improvements in effective area, spatial resolution, and spectral resolution for future instruments. These improvements will revolutionize Solar System studies in the following ways: 
\vspace{-0.5em}
\begin{itemize}
\setlength\itemsep{-0.25em}
\item Investigate early Solar System elemental and molecular composition via comet emissions, including rapid outflow events at the snow line
\item Provide elemental composition maps of the surfaces of satellites throughout the solar system with clear implications for astrobiology at Europa
\item Revolutionise magnetospheric studies of high energy transport and global dynamics of the space environments around planets
\item Study effects of solar X-ray activity on planet atmosphere evolution with broad ramifications for the evolution of exoplanet atmospheres and their star-planet relationships
\end{itemize}
\vspace{-0.25em}
These milestones will usher in a truly transformative era of Solar System science through the study of X-ray emission.}
\vspace*{\fill}

\pagebreak
\setcounter{page}{1}

\section{Solar System Objects as X-ray Sources}

Despite being an area of astronomical research for millenia, the study of Solar System objects has undergone spectacular growth in the past two decades. Recent improvements in satellite technologies coupled with reduced flight costs have generated a swell in interplanetary missions designed for {\it in situ} observations of Solar System objects. Together with frequent, high-resolution observations using Earth-based telescopes and satellites, Solar System objects are monitored at unprecedented spatial and temporal precisions for astronomical systems. The Solar System truly is a perfect laboratory to study astronomical objects and leverage those insights to understand processes in the more distant universe.

Solar System X-ray studies have provided a fundamental understanding of high energy environments in our local universe. Solar System objects emit X-rays through a variety of mechanisms, such as: charge-exchange emissions between ions and neutral particles, fluorescence emission induced by solar X-rays, bremsstrahlung emissions from auroral activity in magnetospheres, and scattering of solar X-rays. These mechanisms have been utilized to study surface and atmospheric compositions \citep{Wargelin2004,  Dennerl2008}, magnetospheric and auroral dynamics \citep{Elsner2005, Bhardwaj2007b, BranduardiRaymont2008}, and energy and mass transport \citep{Gladstone2002, Bodewits2007, Bhardwaj2007, Snios2016}. X-ray observations have provided key insights into: planets (Mercury, Venus, Earth, Mars, Jupiter, Saturn), satellites (Moon, Io, Europa, Ganymede), comets, asteroids (Eros, Itokawa), and space environments (planetary radiation belts, satellite plasma tori, boundary layers such as Earth's magnetopause) \citep[and reference therein]{Bhardwaj2014}. This broad range of emission mechanisms, coupled with the diversity of the physical conditions of the emitting objects, enable Solar System X-ray studies to provide an irreplaceable window on a wide array of astrophysical bodies and environments.  


In looking towards the future of Solar System sciences, interplanetary probe and lander missions (such as those from the {\it Discovery}, {\it New Frontiers}, and {\it Solar System Exploration} programs) will provide invaluable single-point space-time measurements that will radically change our perspective of the local universe. However, it is often challenging to interpret these measurements in a global context. The broader impact for scientific application will come from integrating {\it in situ} measurements with observations from Earth-based telescopes and satellites. Moreover,  simultaneous, remote X-ray instruments are increasingly used to complement {\it in situ} observations. For example, comparisons of remote X-ray observations conjugate with {\it Juno} connect {\it in situ} measurements to the global current systems and particle acceleration processes at giant planets (Gladstone et al., in prep; Rymer et al., in prep). This need for complementary X-ray data to will only continue to grow with the increase of probe and lander missions in the coming decades. 

Beyond the benefits for Solar System science, high-energy observations of planetary atmospheres are vital for exoplanet studies. Elevated stellar X-ray activity may damage atmospheres by removing bulk gas, depleting ozone, and dissociating water \citep{Segura2010}. Alternatively, increases in X-ray activity may beneficially stimulate exotic chemical reactions required for complex molecular compounds, such as amino acids, to form \citep{Glassgold2012, Cleeves2015, Cleeves2017}. Planet atmosphere evolution models rely heavily on the high-precision and sampling rates of Solar System observations. The study of these X-ray-induced interactions is especially valuable for exoplanetary systems with K- and M-class host stars where X-ray flares are frequent and intense. X-ray emissions produced through the rapid rotation of Jupiter's strong magnetic field and plasma injection from Io also provide a spatially-resolved analogue for exoplanets and other rapidly-rotating magnetospheres such as brown dwarfs and pulsars.

While X-ray observations are invaluable for both Solar System and exoplanetary sciences, current X-ray instruments (e.g. \chandra, {\it XMM}) suffer from low effective areas at soft X-ray energies ($\sim$150\,cm$^{2}$) with either limited spatial ($\sim$0.5") or spectral ($\sim$1 E/$\rm \Delta$E) resolutions. High-energy observations of the Solar System therefore need long exposure times, requiring either prohibitively long observing campaigns or extensive modeling of the systems in question in order to probe fundamental physics and processes that occur on timescales shorter than hours. In scenarios where high spatial resolution is required, options are even further limited as only one instrument, \chandra, is presently capable of satisfying such constraints. These restrictions ultimately limit the overall scope and impact for broader scientific application. Incorporating modern X-ray optics will improve effective areas ($>$1000\,cm$^{2}$) and spatial resolutions ($\sim$0.1"), which will answer fundamental questions that are currently present in Solar System sciences. In this paper, we highlight several key scientific advancements in Solar System studies that will be possible through utilizing state-of-the-art X-ray instrumentation and technologies. 

\section{Comets as Tracers of the Early Solar System}

The study of comets has been pivotal in understanding the chemical formation and evolution of the Solar System. Comets are volatile-rich planetesimals created during the initial period of planetary formation. Although the majority of comets currently reside in the outermost regions of the Solar System, orbital perturbations cause a small fraction of these objects to enter into the inner Solar System (< 3 AU) where ices from the comet's surface start to sublimate and are ejected \citep{Meech2004}. This outgassing creates a large cloud of dust and gas that surrounds the comet, known as a coma, as well as generates the comet's iconic tails. The deposition of dust and ice from comets in the inner Solar System, including direct impacts by comets, has been speculated as a potential delivery mechanism of volatile molecules, such as water, to terrestrial planets where such materials are theorized to be lost during initial stages of  planet formation \citep{Anders1989b, Chyba1992, Marty2017}. In studying comets, we therefore obtain detailed information regarding early-stage chemical and dust compositions and the transport of such chemicals throughout the Solar System. 


While widely known as optically bright sources, the past two decades of research have shown comets to also be X-ray bright. The first evidence for comet X-ray emission was found from C/1996 B2 (Hyakutake) by \cite{Lisse1996}, and X-rays have presently been observed from over 30 comets. As a comet approaches perihelion, it generates X-rays from charge-exchange interactions between solar wind ions and neutral gas in the coma \citep{Cravens1997, Wegmann2004, Bodewits2007}. Coherent scattering of solar X-rays by comet dust and ice particles contribute significantly to the total emission intensity at energies greater than 1 keV, providing an additional tracer of the comet's composition \citep{Snios2014, Snios2018}. Cometary X-ray emissions have been used to determine the velocity, composition, and freeze-in temperature of the solar wind \citep{Kharchenko2003, Bodewits2004, Snios2016}, to identify and map plasma interaction structures \citep{Wegmann2005}, and to determine the distribution and composition of the neutral gas \citep{Lisse2005, Mullen2017}.

\begin{SCfigure}
  \caption{An observation of Comet C/2012 S1 (ISON) showcasing optical and X-ray emissions \cite[NASA/CXC]{Snios2016}. The X-ray emission was detected with \chandra{} using a 24-hour exposure during the comet's closest approach with Earth. The notable increase in soft X-ray effective area of future X-ray instruments will allow them to detect the daily fluctuations within comet atmospheres out to a distance 3 times greater than presently possible, while advances in spatial resolution will provide details at resolution of 1000 km.}
  \includegraphics[width=0.69\textwidth]{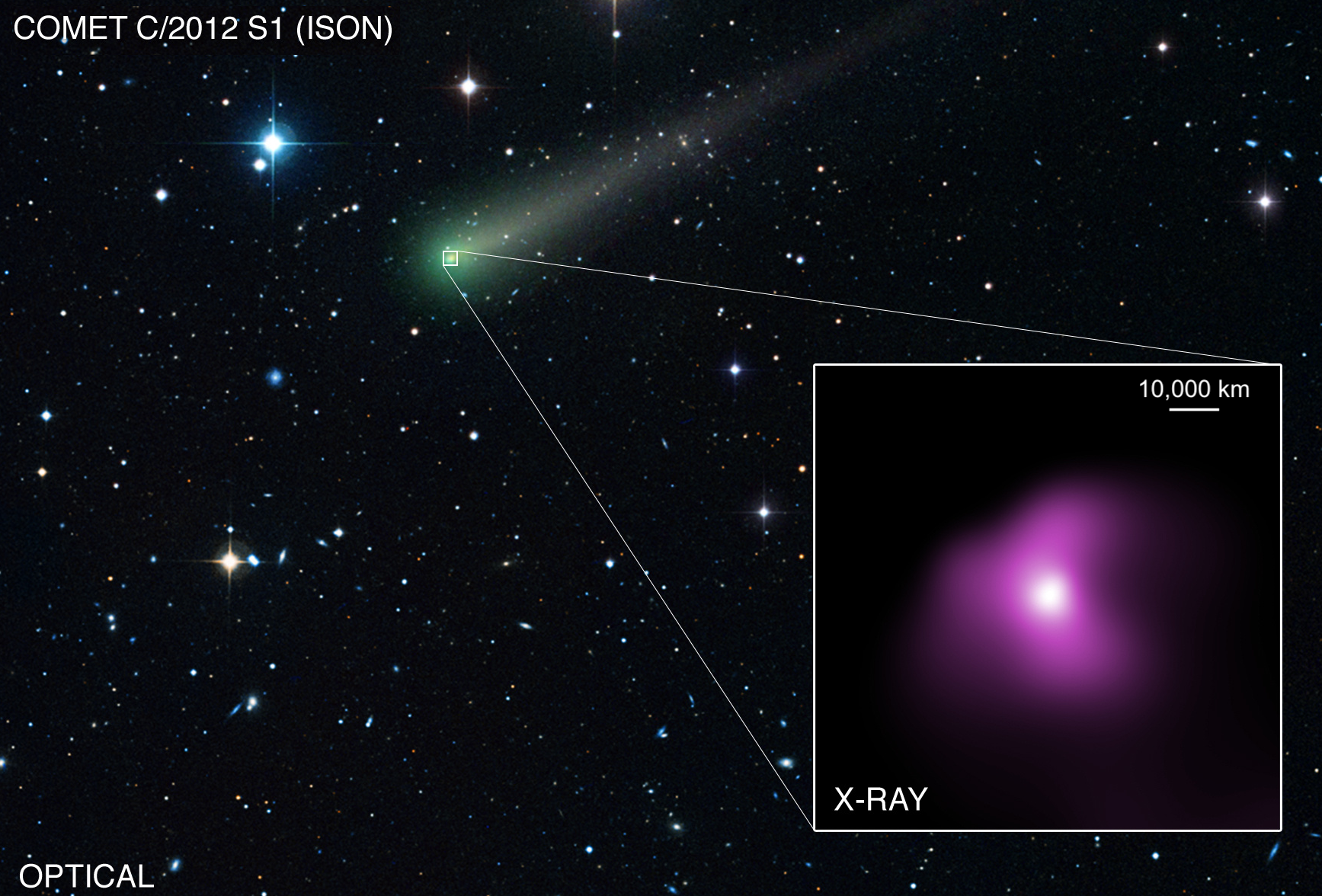}
  \vspace{-1em}
\end{SCfigure}

Limitations of current X-ray instruments restrict observations to near-Earth comets and to the brightest emission periods during perihelion approach. Such restrictions allow for only 1-2 observations per comet at a rate of 0.5 comets per year. Effective area increases to 2 m$^{2}$ at 1 keV will allow future X-ray instruments to observe an average of 4 comets per year. Comet observations will also be viable out to distances of $\sim$2.5 AU, where the outgassing may be driven by CO and CO$_{2}$, allowing for multiple observations throughout perihelion approach and exit. Such fantastic increases in available targets and sampling rates will for the first time allow statistical analyses of cometary X-rays as a function of the comet's composition, outflow rates, origin, orbit, and chemical evolution through its orbit(s). The resulting spatial and temporal statistics from comet observations will additionally provide vital diagnostics for laboratory-based astrochemistry and cross-section studies which presently lack high-precision results for direct comparison \citep[see the white paper by Betancourt-Martinez et al. for further information]{Ali2016, Bodewits2019}. Continued analysis of X-ray emissions from comets is vital for understanding the chemical composition of the early Solar System, probing the evolution and transportation of complex molecular compounds potentially crucial for life, and applying these chemical results for a broader astrophysical use. 

\section{Surface Composition of Planets and Satellites}

Lunar X-rays were first remotely mapped on the Apollo 15 mission \citep{Adler1972}, and through fluorescence line emissions, the locations of oxygen, magnesium, aluminum, and silicon were identified across the sunlit side of the Moon \citep{Schmitt1991, Wargelin2004, Narendranath2011}. Currently, the highest surface resolution is 20 km from the Moon-orbiting {\it Selenological and Engineering Explorer} \citep[{\it SELENE};][]{Yokota2009}. Using a spatial resolution better than the 0.5", future Earth-orbiting X-ray instruments will obtain a 2 km surface resolution for lunar surface abundances. Such improvements will allow accurate extrapolation from the remote map to the narrow field-of-view and high resolution maps of 1-10 km regions performed from lunar orbiters and landers (e.g. {\it Clementine}, {\it Lunar Reconnaissance Orbiter}). Deeper exposures will also allow us to identify abundances for exotic elements, such as gold and sulfur, that are not observable with current instruments. Mapping of the Lunar surface will be vital for future Moon-based mission objectives.

Studies of fluorescence emissions from the surface of Mercury have revolutionised our understanding of the composition of the planet \citep{Weider2015}. Such studies have been possible through {\it in situ} missions such as {\it Messenger}, and in the future {\it Bepicolombo}. While Mercury is likely to be too close to the Sun for X-ray observations by an Earth orbiting observatory, the surface composition of rocky and icy bodies beyond Earth can be studied remotely through fluorescence processes.

In the outer Solar System, Jupiter's moons are constantly bombarded by high energy particles from Jupiter's magnetosphere. These collisions trigger characteristic X-ray fluorescence lines from elements which has allowed \chandra, through virtue of its spatial and spectral resolution, to identify oxygen and sulphur in the surfaces of Io, Europa, and Ganymede \citep[Nulsen et al., in prep]{Elsner2005b}. For Europa in particular, the recent detection of plumes from the sub-surface ocean means that X-ray fluorescence studies of the surface have key implications for identification of ocean compositions. The satellite's spectra also hint at a variety of other emission lines (e.g. sodium), but current sensitivity is insufficient for a clear detection. New instruments with improved sensitivity will map these trace elements, providing key abundances that are critical for understanding the salt content of the sub-surface ocean. These types of observations are currently not possible without the \chandra-like spatial resolution required to distinguish the moons from the background. Improvements in sensitivity may make these studies possible, and extend them to all the Galilean moons at Jupiter and Enceladus at Saturn.

\section{Planetary Magnetospheres and Aurorae}

\begin{figure}[t]
	\begin{tightcenter}
		\includegraphics[width=0.415\textwidth]{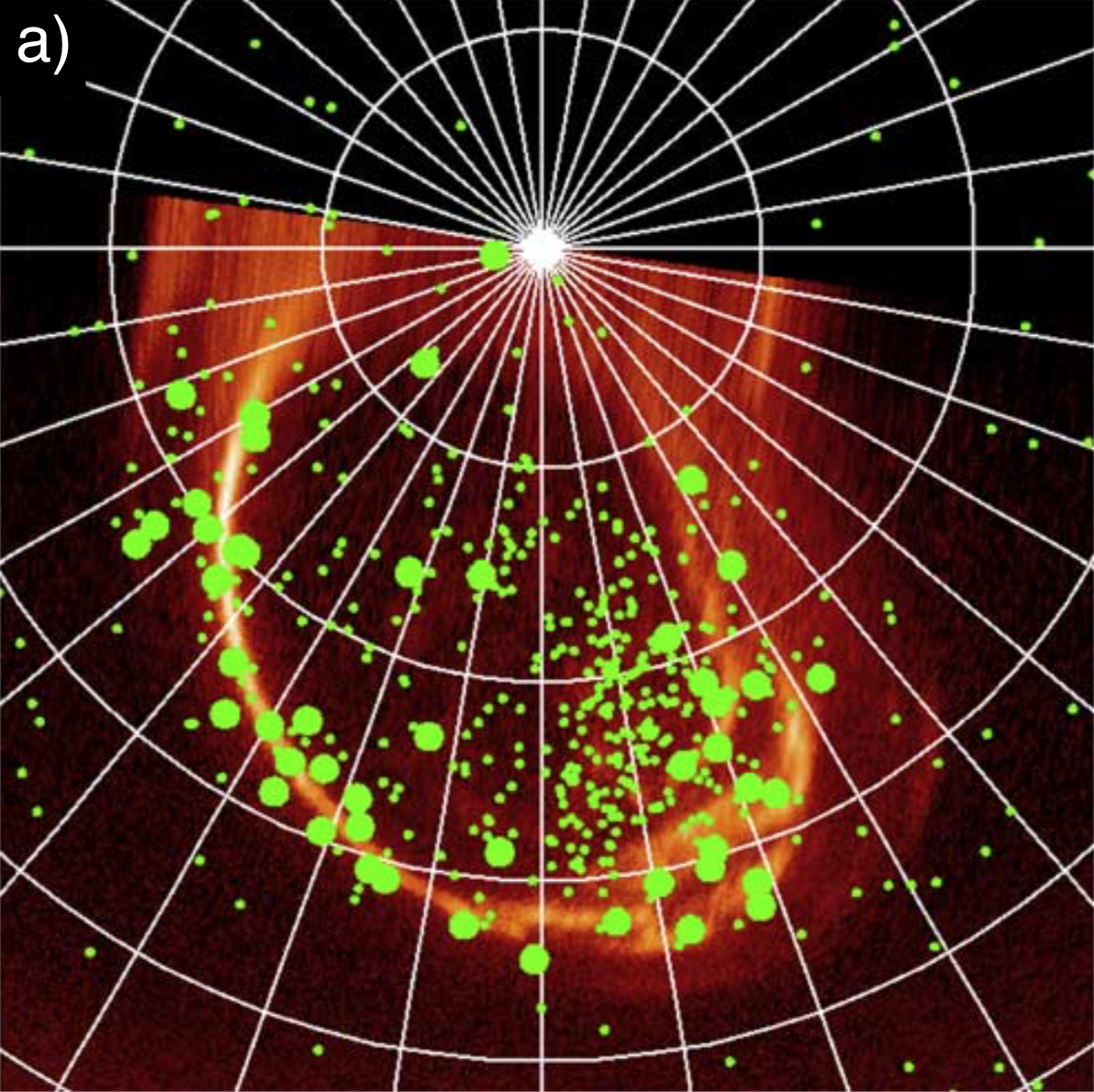}
		\includegraphics[width=0.52\textwidth]{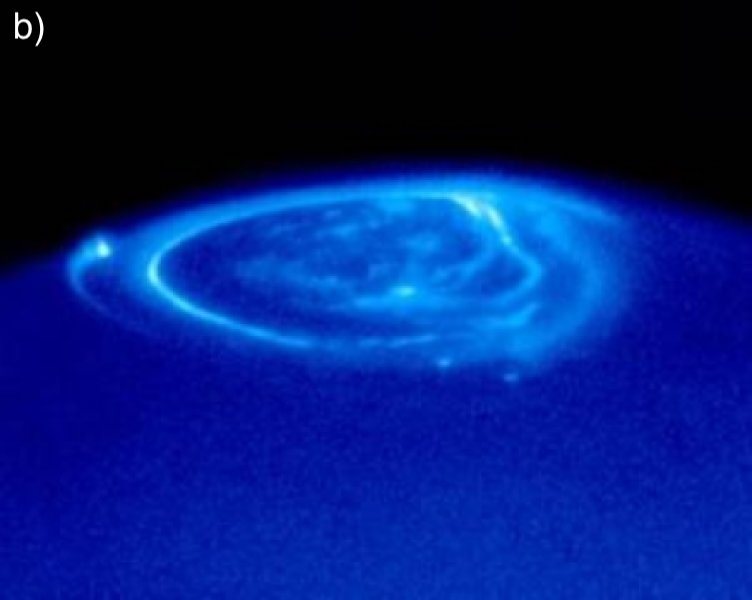}
	\end{tightcenter}
\vspace{-0.5em}
	\caption{a) Overlaid \chandra{} (green dots) and {\it HST} (orange) polar projection of Jupiter's North Pole \citep{BranduardiRaymont2008}. Large green dots indicate hard X-ray bremsstrahlung emissions from precipitating electrons on current systems outwards from the planet. Small green dots indicate X-rays from Jupiter's returning currents, through which precipitating high energy ions charge exchange with hydrogen in Jupiter's atmosphere. b) A UV auroral image showing hydrogen excitation emission from electron collisions in the atmosphere \citep{Clarke2004}. While the UV auroral emissions have permitted a broad range of studies of Jupiter's global dynamics through the upward current system, poor sensitivity on previous X-ray instruments has limited interpretation of the partner returning current system. Advances in X-ray sensitivity will allow X-ray studies to provide comparable scientific impact to the UV observations. }
\vspace{-1em}
\end{figure}

This year marks the 40th anniversary of the discovery that planets other than Earth generate X-rays, with the discovery of Jupiter's bright and dynamic X-ray aurorae in 1979 \citep{Metzger1983}. Since this novel finding, X-ray observations by the {\it Einstein}, {\it ROSAT}, \chandra, {\it XMM-Newton}, {\it Suzaku}, and {\it NuSTAR} X-ray observatories have all provided deep insights into the planetary properties of Jupiter, its moons, its surrounding space environment, and its interaction with the upstream solar wind. Jupiter provides a critical local analogue for more distant gas giants and rapidly-rotating magnetospheres such as exoplanets, brown dwarfs and pulsars, for which there are no {\it in situ} measurements with which to enrich understanding of remote observations.

\chandra's higher spatial resolution has been fundamental in understanding the nature of these processes. Within one year of launch, \chandra{} revealed that the majority of  Jupiter's X-ray aurorae were localised in a region at Jupiter's poles \citep[see also Figure 2a]{Gladstone2002} in which the planet acts as the largest natural particle accelerator in the Solar System \citep{Barbosa1990}, accelerating ions to tens of MeV energies. \chandra's spatial resolution has further revealed a variety of fundamental properties of Jupiter, showing that Jupiter's poles sometimes pulse with regularity and can operate independently of one another \citep{Gladstone2002, Dunn2017} and helping to reveal how the giant planets respond to space weather events \citep{Dunn2016}. 

The ion precipitations that predominantly produce Jupiter's X-ray emission reveal the dynamics of Jupiter's return current \citep{Cravens2003}, which can only be remotely studied through X-ray observations. While the X-rays uniquely permit studies of the return currents, the current system outwards from the planet can be probed by both UV and X-ray observations \citep[Figure 2a]{BranduardiRaymont2008}. Over the last two decades, UV observations of Jupiter's outward current system (Figure 2b) have produced hundreds of findings and profoundly changed our understanding of giant planets (e.g. review in \citealt{Grodent2015}). It is not yet possible to leverage the X-rays in the same manner, due to the low instrument sensitivities and consequent low statistics. For instance, Figure 2a shows several hours of \chandra{} observations which attained a few hundred photons. Improvements in sensitivity will provide $\sim10^4$ counts for a similar duration. This transformation, in combination with high angular resolution, will identify the auroral morphology of the return currents on the few minute-timescales in which they are known to vary \citep{Dunn2017, Jackman2018}. This will allow X-ray observations to probe the return currents, as UV observations have done for outward currents, opening revolutionary new insights into the global dynamics of rapidly-rotating magnetospheres and giant planets.

Beyond the Jovian system, X-ray observations of Saturn have revealed atmospheric properties \citep{Ness2004, BranduardiRaymont2010} and fluorescent emissions from the rings, identifying elemental composition \citep{Bhardwaj2005}. However, despite Saturn's comparable size and rotation-rate to Jupiter, these studies have failed to detect X-ray aurora at the planet. This raises the question of whether Jupiter and Saturn are fundamentally different in their particle acceleration or whether it is simply that Saturn's X-ray aurorae are too dim to detect with current instrument effective areas. Future X-ray instruments will determine whether X-ray aurorae are ubiquitous across gas giants or dependent upon specific properties of each system. A higher sensitivity instrument will also open up observations of the ice giants, Uranus and Neptune, to identify the high energy environments around these planets. 


An instrument with heightened sensitivity and a comparable spatial resolution to \chandra{} will revolutionise the study of planetary magnetospheres, providing a new window into the high-energy regimes of these planets.

\section{Summary}
The Solar System is a perfect laboratory to study astronomical objects at remarkable spatial and temporal precisions and to leverage those insights to understand the more distant universe. X-ray instruments are fundamental for such studies by providing insights into Sun-object interactions, developing surface composition maps, probing small-scale and global magnetospheric dynamics, and tracking local astrochemical reactions. Although current X-ray instruments are still a vital tool to augment planetary science in the next decade, their limitations mean that a giant leap in this science can only be achieved with a major step up in observatory capabilities. Implementing modern X-ray optics in future instruments, including improvements in effective area, spatial resolution, and spectral resolution, will remove these strict limitations and foster a truly transformative era of Solar System science through the study of X-ray emission.


\pagebreak
\section*{References}
\begin{multicols}{2}
\begingroup
\renewcommand{\section}[2]{}%
\bibliographystyle{aasjournal_mod}
\bibliography{all_data}
\endgroup
\end{multicols}
\end{document}